\documentclass{ceurart}

\usepackage[T1]{fontenc}
\usepackage[utf8]{inputenc}
\usepackage[normalem]{ulem}
\usepackage[]{outlines}
\usepackage[dvipsnames]{xcolor}
\usepackage{mathpartir}
\usepackage{iftex}
\ifluatex
  \usepackage{lua-visual-debug}
\else
\fi
\usepackage{float}
\usepackage{placeins}
\usepackage{svg}
\usepackage{amsmath,amssymb,comment}
\usepackage{listings}
\PassOptionsToPackage{hyphens}{url}
\usepackage{hyperref}
\hypersetup{                    
  unicode,
  bookmarksnumbered,
  colorlinks,
  pdfauthor={Hader and Ozdemir},
  pdftitle={An SMT-LIB Theory of Finite Fields},
  pdfborder={0 0 0},
  pdfmenubar=false,
  pdfpagemode=UseNone,
  pdfnonfullscreenpagemode=UseNone,
  bookmarksopen=false,
  breaklinks=true,
  linkcolor={green!50!black},
  citecolor={red!50!black},
  urlcolor={blue!50!black}
}

\usepackage{xspace}
\usepackage{nicefrac}
\usepackage{tikz}
\usetikzlibrary{arrows.meta}
\tikzset{>=Latex[]}
\usepackage{booktabs}
\usepackage{menukeys}
\usepackage{caption}
\usepackage{subcaption}
\usepackage[linesnumbered,noend]{algorithm2e} 
\usepackage[inline]{enumitem}

\usepackage{mathtools}

\DeclarePairedDelimiter\floor{\lfloor}{\rfloor}

\newcommand{\xspacemm}{\ifmmode\else\xspace\fi}
\newcommand{\mathnoun}[2]{\newcommand{#1}{\ensuremath{#2}\xspacemm}}

\mathnoun{\C}{\mathcal{C}}
\mathnoun{\FF}{\mathbb{F}}
\mathnoun{\F}{\mathbb{F}}
\mathnoun{\ZZ}{\mathbb{Z}}
\mathnoun{\GG}{\mathbb{G}}
\mathnoun{\NN}{\mathbb{N}}
\mathnoun{\smod}{\mathsf{smod}}

\newcommand{\smtlib}{SMT-LIB\xspace}
\newcommand{\cdclt}{CDCL($\mathcal{T}$)\xspace}

\newcommand{\slc}[1]{\mbox{\upshape\texttt{#1}}}
\newcommand{\ffsort}[2]{\slc{(\_ FiniteField #1\xspace #2)}\xspace}
\newcommand{\ffsortp}{\ffsort{$p$}{}}
\newcommand{\ffsortpn}{\ffsort{$p$}{$n$}}

\newcommand{\ffop}[1]{\texttt{(#1 \ffsortp \ffsortp )}\xspace}
\newcommand{\ffopp}[1]{\texttt{(#1 \ffsortp \ffsortp \ffsortp )}\xspace}
\newcommand{\ffoppassoc}[1]{\texttt{(#1 \ffsortp \ffsortp \ffsortp\ :left-assoc)}\xspace}

\newcommand{\gbs}{Gr\"obner bases\xspace}

\newtheorem{example}{Example}

\begin{document}

\copyrightyear{2024}
\copyrightclause{Copyright for this paper by its authors.
	Use permitted under Creative Commons License Attribution 4.0
	International (CC BY 4.0).}

\conference{SMT 2024: Satisfiability Modulo Theories,
	July 22--23, 2024, Montreal, Canada}

\title{%
  An SMT-LIB Theory of Finite Fields
}

\author[1]{Thomas Hader}[%
email=thomas.hader@tuwien.ac.at,
]
\author[2]{Alex Ozdemir}[%
email=aozdemir@cs.stanford.edu,
]

\address[1]{TU Wien,
	Favoritenstraße 9-11, 1040 Wien,
	Austria}

\address[2]{Stanford University;
	353 Jane Stanford Way; Stanford,
	CA, 94305 USA}


\begin{keywords}
	SMT \sep
	finite fields \sep
	non-linear reasoning
\end{keywords}

\maketitle

\begin{abstract}
In the last few years there have been rapid developments in
SMT solving for finite fields.
These include
new decision procedures,
new implementations of SMT theory solvers,
and
new software verifiers that rely on SMT solving for finite fields.
To support interoperability in this emerging ecosystem,
we propose the SMT-LIB theory of finite field arithmetic (FFA).
The theory defines a canonical representation of finite field elements
as well as
definitions of operations and predicates on finite field elements.
\end{abstract}


\section{Introduction}\label{sec:intro}
Finite fields are the basis for a large body of
security-critical code.
They are used in public-key cryptography:
elliptic curves over finite fields are used in nearly all web browser connections
for key exchange or digital signatures~\cite{nist_ec,longitudintaltls,anderson2019tls}.
They are used in private-key cryptography:
in both the Poly1305 message authentication code~\cite{bernstein2005poly1305}
and Galois counter mode (GCM)~\cite{salowey2008rfc}.
They are also the basis of most protocols for secure
computation.
For instance,
many zero-knowledge proof systems
prove and verify predicates expressed as finite field
equations~\cite{STOC:GolMicRac85,parno2016pinocchio,groth2016size,zkp_vuln_sok}.
Also,
many secure multi-party computation protocols
evaluate circuits over finite fields~\cite{damgaard2012multiparty,mpc_sok}.
Finally, some homomorphic encryption schemes apply to data in a finite
field~\cite{regev2009lattices,fhe_sok}.

The importance and prevalence of (finite-)field-based programs
creates a need for tools that can formally verify them.
Ideally, such tools would be partially or fully automated.
The natural approach is \textit{SMT-based verification},
as taken by prior tools like Dafny~\cite{leino2010dafny} and
Boogie~\cite{barnett2006boogie}.
In this approach, a software \textit{verifier}
reduces the correctness of the
program to logical
formulas which it dispatches to
a \textit{satisfiability modulo theories} (SMT) solver.
Applying this approach to field-based software
generally requires an SMT solver
that can solve finite field equations.

One way to solve field equations is by encoding them as integer equations,
which many SMT solvers already comprehend.
Consider (for the moment) a finite field of prime order $p$.
Such a field is isomorphic to the integers $\{0, \dots, p-1\}$
with addition and multiplication modulo $p$~\cite{dummit2004abstract}.
Thus, (non-linear) equations mod $p$ can be encoded as (non-linear)
integer equations.
In this encoding, an equation $xy=z$ over field variables $x, y, z$
would be encoded as $(x'y'-z') \bmod p =0$, where $x',y',z'$ are the integer
representatives of the field variables.
These equations can now be solved using an integer solver.
Or, since all terms can be bounded, they can be solved as bit-vector (bounded integer) equations.
However, prior experiments have shown that existing integer and bit-vector
solvers perform poorly when given inputs that encode finite field
arithmetic~\cite{ozdemir2023satisfiability,bb_abstractions}.

To overcome the limitations of encoding,
two direct SMT theory solvers for finite fields have recently emerged.
The first is an MCSat~\cite{jovanovic2013design} solver
that is implemented in
Yices~\cite{hadermsthesis,hadersmt22,LPAR:HKK23,dutertre2014yices,hader2024mcsatbased}.
The second is a CDCL(T) solver
that is implemented in cvc5~\cite{ozdemir2023satisfiability,split_gb,cvc5}.
%
%
Currently, these solvers accept field terms and equations expressed using a
bespoke extension to SMT-LIB.
This extension has not been standardized.

These SMT solvers have already enabled a variety of research projects and tools for
automatically
verifying systems that use zero-knowledge proofs (ZKPs).
One project builds an automatically verifiable compiler pass for CirC:
a compiler used with ZKPs~\cite{ffblast}.
Another builds a tool $\text{QED}^2$ that automatically verifies ZKP code in the
Circom language~\cite{PLDI:PCWRVMCGFD23,circom}.
Another builds a tool for automatically verifying ZKP code written using the
Halo2 library~\cite{halo2ver}.
All of these projects use an SMT solver with finite field support.

Given the long-term importance of finite fields to security-critical software,
the emergence of multiple SMT solvers with finite field support,
and
the emergence of multiple automatic verification tools expecting finite field
support,
we think the time is ripe to specify finite fields as an SMT-LIB theory.
In this short paper, we do exactly that.
In our specification, we consider all finite fields:
those of prime order and their extensions.
We consider fields of arbitrary size.
Many cryptosystems require large fields
(such as a prime order field with $p \approx 2^{256}$
or the binary extension field of order $2^{128}$),
but some can also operate over smaller fields
(such as 32-bit or 64-bit fields)
\unskip~\cite{plonky2,ben2018fast}.

\paragraph{Related Work}
There is already much work on
verifying cryptographic implementations
through \textit{interactive} theorem proving
and verification languages.
Examples of secret-key and public-key cryptography include
  Fiat cryptography~\cite{fiat_crypto},
  Easycrypt~\cite{barthe2012easycrypt},
  HACL*~\cite{zinzindohoue2017hacl},
  and
  Jasmin~\cite{almeida2017jasmin}.
There is also some work on interactive verification
for ZKPs
in the context of the Leo compiler~\cite{kestrel},
and by using refinement proofs~\cite{liu2023certifying,jiang2023less}.
With better SMT-level support for finite fields,
ITP proof automation for finite field lemmas could be improved through
ITP-SMT bridges, like SMTCoq~\cite{ekici2017smtcoq}.

Further afield, some cryptographic implementations have been modeled and analyzed
using \textit{automated} symbolic analysis tools like
Tamarin~\cite{meier2013tamarin} and
ProVerif~\cite{blanchet2018proverif}.
The benefit of these tools is their high level of interpretability and automation,
which allows them to be applied to protocols of realistic complexity,
such as Signal~\cite{cohn2020formal}.
However, they struggle to accurately
model algebraic structures~\cite{cremers2019prime}.
SMT-level algebraic reasoning would complement this research.

Another line of research develops SMT solvers
for
non-linear integer and real arithmetic
using
CDCL(T)
\unskip~\cite{abraham2021deciding,
cimatti2018experimenting,
marechal2016polyhedral,
franzle2006efficient,
tung2016rasat,
jovanovic2011cutting,
dillig2009cuts,
corzilius2015smt}
and
MCSat
\unskip~\cite{
jovanovic2013solving,
jovanovic2017solving,
moura2013model}.
Some works specifically consider local search
\unskip~\cite{
cai2023local,
zhang2024deep,
wang2023efficient}
and quantifier elimination
\unskip~\cite{
caviness2012quantifier,
weispfenning1997quantifier}.
This research serves as good inspiration into techniques for finite field
solving.

\section{Background}\label{sec:bg}
We provide a brief summary of the relevant concepts of finite fields.
A comprehensive overview can be found in~\cite{lidl1994introff,
mceliece2012finite, contemporaryAbstractAlgebra} as well as in many other
algebra textbooks.

\paragraph{Fields.}
A field $\F$ consists of a set of elements $S$ on which the two binary operators
\emph{addition} ``$+$'' and \emph{multiplication} ``$\cdot$'' are defined.
$ S $ is closed under both operators, i.e. when applied on two elements of $S$, the result is in $S$.
Both operators are commutative, associative, and have distinct neutral elements (denoted as \emph{zero} (0) and \emph{one} (1), respectively).
Each element in $S$ has an additive inverse element and all elements in $S\setminus \{0\}$ have a multiplicative inverse element.
Further, distributivity of multiplication over addition holds.
Informally speaking, a field is a set with well-defined operations addition, subtraction, multiplication, and division (with the exception of division by $0$).
Well known examples of fields are the rational number $\mathbb{Q}$ and the real numbers $\mathbb{R}$.

\paragraph{Finite Fields.}
A field \F where $S$ is finite is called a \textit{finite field}.%
\footnote{In honor of French mathematician \'Evariste Galois, finite fields are also called \emph{Galois fields}.}
The size of $S$ is the \textit{order} of \F.
It has been shown that every finite field has order $q$ that is a prime power $q = p^n$.
We distinguish between \emph{prime fields} with $n = 1$ and \emph{extension fields} with $n > 1$.
All fields of equal order are isomorphic
(i.e. equivalent up to relabelling of elements),
thus the field of order $q$ is unique (up to isomorphism).

\paragraph{Prime Fields.}
The prime field of order $p$
can be represented as $S = \{-\floor*{\frac{p-1}{2}},\dots,0,1,\dots,\floor*{\frac{p}{2}}\}$%
\footnote{
  In the (isomorphic) representation $S = \{0, \dots, p-1\}$,
  (with addition and multiplication modulo $p$),
  small ``negative'' values (such as $-1$) are instead
  represented as large positive values (such as $p-1$),
  which can be unintuitive to read.
  We choose our representation because small negative values are common
  in many applications.
}
and is denoted $\F_p$.
Let the function $\smod_p: \ZZ \to S$ be defined
to map $z \in \ZZ$ to the unique element of $S$ that is equivalent to $z$,
modulo $p$.
The function is called ``$\smod_p$'' because it outputs a \textit{signed}
representation of its input.

Addition and multiplication on $S$ are defined by the usual integer operations
followed by an application of $\smod_p$.
Due to the construction of $S$, finding the additive inverse is as simple as flipping the sign (assuming odd $p$).
\begin{example}
	The finite field $\F_5$ can be represented by the integers $\{-2,-1,0,1,2\}$.
	%
	In this representation of $\F_5$, $2 + 1 = -2$, $2 \cdot (-1) = -2$, and $ (2 + 1) \cdot 2 = 1$ hold.
\end{example}

\paragraph{Extension Fields.}
Let $\F_p[\alpha]$ be the set of univariate polynomials
in variable $\alpha$ with coefficients from $\F_p$,
and let $f \in \F_p[\alpha]$ have degree $n$
and be irreducible (i.e.\ it cannot be represented as the product of two non-constant polynomials).
The extension field of order $p^n$
is denoted $\FF_{p^n}$
and can be represented as polynomials in $\F_p[\alpha]$ of degree less
than $n$, with (polynomial) addition and multiplication modulo $f$.
Note that, in this representation, $\{0, 1\} \subseteq \F_p\subseteq\FF_{p^n}$.

\begin{example}\label{ex:extensionField}
	The finite field $\F_{3^2}$ is represented by the following polynomials of $\F_3[\alpha]$ modulo $\alpha^2 - \alpha - 1$:
	\[\{
	0,\,
	\alpha,\,
	\alpha + 1,\,
	-\alpha + 1,\,
	-1,\,
	-\alpha,\,
	-\alpha - 1,\,
	\alpha - 1,\,
	1
	\}\]
	Over $\F_{3^2}$ it holds that $(\alpha + 1) \cdot \alpha = (-\alpha + 1)$.
\end{example}
As the choice of $f$ is not unique in general, different (isomorphic) representations of $\FF_{p^n}$ exist,
even if the representation of $\FF_p$ is fixed.
Note that no finite field
is an \textit{ordered field}.
That is, there is no total ordering
on $S$ that is compatible with the field operations.

\paragraph{Conway Polynomials}
Algebraic tools have many choices for how to represent fields internally.
But, to facilitate interoperability,
the computer algebra community has agreed upon a canonical family of irreducible
polynomials that should be used to represent elements of an extension field $\FF_{p^n}$ in tool interfaces.
These are called the \textit{Conway polynomials} $C_{p,n} \in
\FF_p[\alpha]$,
where $p$ is a prime and $n>1$.
The precise definition of the Conway polynomials is not important for our purposes.%
\footnote{
  The Conway polynomial $C_{p,n}$ is the lexicographically minimal monic
  primitive polynomial of degree $n$ over $\F_p$ that is compatible with
  $C_{p,m}$ for all $m$ dividing $n$.
  Let $r = (p^n-1)/(p^m - 1)$ (which is an integer).
  Then,
  $C_{p,n} \in \FF[\alpha]$ is compatible with
  $C_{p,m} \in \FF[\alpha]$
  if
  for every root $\alpha_0\in\FF_{p^n}$
  of the former,
  $\alpha_0^r$ is a root of the latter.
  The lexicographic ordering used is also slightly non-standard.
  Define the alternating-sign coefficient representation of polynomial $f \in \FF[\alpha]$ 
  to be
  $f = \sum_{i=0}^d (-1)^ic_{d-i}\alpha^{d-i}
  =
  c_d\alpha^d-c_{d-1}\alpha^{d-1}+\cdots+(-1)^da_0$,
  with $c_i \in \{0, \dots, p-1\}$.
  Then,
  the order is lexicographic over the tuples $(c_d, \dots, c_0)$.
}
There is an algorithm for finding them~\cite{heath2004conway}
and they have been pre-computed for many $n$ and $p$~\cite{conway_poly_db}.
The Conway polynomials are used by all prominent computer algebra libraries:
Sage, Magma, GAP, Singular, etc.
We will use the Conway polynomials to define an SMT-LIB syntax for extension field element
literals (Sec.~\ref{sec:theory:lit}).

\begin{example}%
  \label{ex:conway}
  The Conway polynomial $C_{3,2}$ is $\alpha^2-\alpha-1$,
  which is the irreducible used to represent $\FF_9$ in
  Example~\ref{ex:extensionField}.
\end{example}

\section{A Theory of Finite Fields}\label{sec:theory}
This section presents the \smtlib (version 2.6)
\emph{theory of finite field arithmetic} (FFA).
Based on the theory of finite field arithmetic
are the logics of
quantifier-free finite field arithmetic \texttt{QF\_FFA}
as well as its quantified version \texttt{FFA}.

\subsection{The Finite Field Sorts}
\label{sec:theory:sort}

The theory of finite fields defines two kinds of finite field sorts, \emph{prime
field sorts} and \emph{extension field sorts} for prime and extension fields,
respectively (Sec.~\ref{sec:bg}).
They are represented by an indexed sort identifier of the form \ffsortp and
\ffsortpn for prime and extension field sorts, respectively.
The indexes $p$ and $n$ are numerals specifying the finite field order $q =
p^n$.
The index $p$ must be a prime number in both cases.
For extension field sorts, $n > 1$ must hold, as otherwise the resulting sort
would be a prime field.
Providing a non-prime number as $p$ may result in unspecified solver behavior, although solvers are encouraged to report an error.
\unskip\footnote{
  We recommend that solvers test $p$'s primality probabilistically,
  for example
  with a 40-repetition
  Miller-Rabin test~\cite{rabin1980probabilistic}.
  If $p$ is not prime, the solver can report an error.
  If $p$ is prime or if the test is inconclusive,
  the solver may assume that $p$ is prime and continue.
}
As is usual for an indexed sort, two finite field sorts with a different order are different sorts.
Solvers implementing this theory are not required to support extension field
sorts and may report an error in case an extension field sort is specified.
\unskip\footnote{
  We choose different syntaxes for prime fields and their extensions
  so that a user who is only interested in prime fields
  need not understand or even be aware of extension fields.
}

For the rest of this chapter, a \emph{finite field sort} is a prime or extension field sort with an arbitrary fixed order.

\begin{example}\label{ex:define}
  Set the logic to non-linear finite field arithmetic and define finite field sorts of size $5$ and $9$:
  \begin{lstlisting}
  (set-logic QF_FFA)
  (define-sort FF5 () (_ FiniteField 5))
  (define-sort FF9 () (_ FiniteField 3 2))
  \end{lstlisting}
\end{example}

\subsection{The Domain of Finite Field elements}
\label{sec:theory:domain}
A finite field of a given order is uniquely defined up to isomorphism.
Thus, for the sake of defining an SMT theory for finite fields, a canonical representation for a finite field of a given order needs to be fixed.
Otherwise different solvers might present the same model differently.

For a prime field with prime order $p$, the elements are represented by the integers of the set $\{-\floor*{\frac{p-1}{2}},\dots,\floor*{\frac{p}{2}}\}$.
Operations are performed with regard to the function $\smod_p$ as defined in Section~\ref{sec:bg}.
For an extension field of order $p^n$ the field elements are represented by univariate polynomials over the prime field of order $p$.
The implied field is $\FF[\alpha]/C_{p,n}$,
where $C_{p,n}$ is the Conway polynomial (Sec.~\ref{sec:bg}).
All polynomial operations are performed modulo the Conway polynomial $C_{p,n}$.

\subsection{Finite Field Literals}
\label{sec:theory:lit}
In the theory of finite fields, each element of a finite field sort is
represented by a literal.
To avoid confusion with the theories of integer and reals, finite field literals are prefixed with the string \slc{ff}.
We further say that a literal is \emph{normalized} when it stands for an element from the fixed field representation as defined in Section~\ref{sec:theory:domain}.
Non-normalized literals are allowed as an input and are mapped to the
corresponding normalized literal, however, solvers are required to resort to
normalized literals when printing a value.

\paragraph{Prime field literals.}
As stated in Section~\ref{sec:bg}, elements of prime fields can be represented
as integers modulo the field size.
This property is used to define literals in the form of \slc{ff$N$}, where $N$
is an integer value.
Given a prime field sort \ffsortp with prime order $p$, elements represented by
the literals \slc{ff$N$} for all values $N \in
\{-\floor*{\frac{p-1}{2}},\dots,\floor*{\frac{p}{2}}\}$ are normalized.
Every literal outside this set is mapped to the corresponding normalized literal
by utilizing $\smod_p(N)$.
Using this operation, the input gets mapped to the (unique) normalized
representative of the same congruence class modulo $p$.
In the presence of finite field theory,
the user may not define their own symbols
of form \slc{ff$N$}
(nor may they shadow other theory-defined symbols).

\begin{example}
  Given the defined sorts of Example~\ref{ex:define},
  then examples of normalized elements of \lstinline|FF5| are
  \lstinline|ff1|, \lstinline|ff0|, and \lstinline|ff-2|.
  The (non-normalized) literals \lstinline|ff4| and \lstinline|ff10| describe the same element as the normalized literals \lstinline|ff-1| and \lstinline|ff0|, respectively.
\end{example}

\paragraph{Extension field literals.}
Elements of extension field sorts are polynomials.
Literals representing elements of extension field sorts are uniquely describing polynomials by specifying their coefficients.
As described in Section~\ref{sec:bg}, an element of a finite field of order $q = p^n$ with $n>1$ is a polynomial $P \in \F_p[\alpha]$ of degree at most $n-1$.
Let $P = c_0 + c_1\alpha^1 + \cdots + c_{n-1}\alpha^{n-1}$ where $c_i \in \mathbb{F}_p$, then the element $P$ is represented by the literal \slc{ff$c_0$.$c_1$.$\cdots$.$c_{n-1}$} where all $c_i$ are integers.
Tailing zeros may be omitted, e.g.\ in \ffsort{3}{6} the literals \slc{ff1.0.-1.0.0} and \slc{ff1.0.-1} both represent the element $1 - \alpha^2$.
This further ensures that elements in $\F_p\subseteq\F_{p^n}$ have the same literal representation in both \ffsortp and \ffsortpn.
A (polynomial) literal of \ffsortpn is normalized when all integer coefficients are normalized with regard to $p$ and all tailing zeros are omitted.
Specifying a literal with more than $n$ coefficients is invalid.

\begin{example}
  Again, given the defined sorts of Example~\ref{ex:define},
  normalized literals of \lstinline|FF9| are all (normalized) literals of \ffsort{3}{}:
  \lstinline|ff0|, \lstinline|ff1|, and \lstinline|ff-1|, as well as literals representing the further (polynomial) elements of $\F_{3^2}$,
  for example \lstinline|ff-1.1|, \lstinline|ff0.1|, and \lstinline|ff-1.-1| representing the elements
  $\alpha - 1$, $\alpha$, and $-\alpha - 1$, respectively.
  Note that \lstinline|ff2.1| and \lstinline|ff1.0| are both non-normalized versions of \lstinline|ff-1.1| and \lstinline|ff1|, respectively.
\end{example}

\paragraph{Well-Sortedness of literals.}
Since the defined literals are overloaded (for example, every finite field has a one element, thus \texttt{ff1} is of undetermined sort), the order of the literal's field must be specified in order to satisfy the well-sortedness requirements of \smtlib.
There are two ways of specifying the sort of a finite field literal.
\begin{enumerate*}[label=(\roman*)]
\item
By indexing the literal \slc{(\_ ff$\dots$ $p$)} and \slc{(\_ ff$\dots$ $p$ $n$)} with the finite field order $p$ and $p^n$, respectively.
This allows the literal's sort to be derived, as the order of the finite field specifies the sort uniquely.
%
%
\item
Since $p$ might be a large number, a short-cut is provided by \slc{(as ff$\dots$ $S$)} where $S$ is a finite field sort.
Using the \slc{define-sort} command, one can assign a symbol to a finite field sort to be used for all its literals.
\end{enumerate*}
Table~\ref{tab:syntax:lit} gives an overview of all variants.

\begin{table}[t]
  \centering
  \begin{tabular}{llll}
    \toprule
    field type  & order & syntax & syntax type \\
    \midrule
    prime field & $p$ & \slc{(\_ ff$N$ $p$)} & indexed \\
                && \slc{(as ff$N$ \ffsortp)} & annotated \\
    \midrule
    extension field & $p^n$ & \slc{(\_ ff$N$.$\cdots$.$N$ $p$ $n$)} & indexed\\
                    && \texttt{(as ff$N$.$\cdots$.$N$ \ffsortpn)} & annotated \\
    \bottomrule
  \end{tabular}
  \caption{
    Different ways to define literals.
  }
  \label{tab:syntax:lit}
\end{table}

\begin{example}\label{ex:label}
  The expressions \lstinline|(_ ff1 5)| and \lstinline|(_ ff1 3 2)|
  denote the multiplicative identity (one) of $\F_5$ and $\F_{3^2}$, respectively.
  %
  Using the defined sorts from Example~\ref{ex:define}, \lstinline|(as ff1 FF5)| and \lstinline|(as ff1 FF9)| can be used alternatively.
\end{example}

\paragraph{Reporting Models.}
To make it easy for tools to parse models,
all solvers must report models using normalized
literals in the indexed representation.

\subsection{Finite Field Operations}
\label{sec:theory:ops}

All of the following operator definitions represent well known semantics from algebra.
Thus, an explicit definition of their semantics is omitted and we refer to Section~\ref{sec:bg} for further details.
Operations always operate on one specific finite field sort $ S $, i.e.\ all parameters have sort $S$ and an element of $S$ is returned.
All functions are defined for all prime field sorts as well as all extension field sorts.
For the sake of brevity, the extension field sort variants of the functions are omitted.
Table~\ref{tab:syntax:ops} gives an overview of all operations.

\paragraph{Binary arithmetic.}
For each finite field order, we define operations that take two finite field elements of one finite field sort and return an element of the same sort.
Given two inputs, the operations represent sum, difference, product, and quotient, respectively.%
\footnote{
  Since \texttt{ff.neg} and \texttt{ff.recip} are defined as well,
  \texttt{ff.sub} and \texttt{ff.div} are redundant.
  However, we believe that all common mathematical operations
  should have operations in the finite field theory.
  Furthermore, other arithmetic theories in \smtlib
  also define redundant subtraction and division operators.
}
\begin{align*}
&\ffoppassoc{ff.add}\\
&\ffopp{ff.sub}\\
&\ffoppassoc{ff.mul}\\
&\ffopp{ff.div}\\
\end{align*}

As hinted by the \texttt{:left-assoc} keyword, occurrences of \texttt{ff.add} and \texttt{ff.mul} may contain more than two arguments and multiple arguments are grouped left associatively.
However, note that both operations are associative anyway.
\unskip\footnote{
  Note that \texttt{ff.div} is not Euclidean division, rather it is multiplication
  by an inverse in the field.
  Thus, the remainder of a division, i.e. \texttt{ff.rem}, would not be meaningful.
}

\paragraph{Unary arithmetic.}
For each finite field sort, there are the following unary operations:
\begin{center}
  \ffop{ff.neg}\\
  \ffop{ff.recip}\\
\end{center}

Here, \lstinline|ff.neg| returns the unary negation (usually written as $-x$ for an element $x$), i.e.\ the inverse element with regard to addition.
The operation \lstinline|ff.recip| returns the reciprocal value (usually written as $x^{-1}$ for an element $x$), i.e.\ the inverse element with regard to multiplication.

\paragraph{Division by zero.}
Two operators
(\texttt{ff.div}, \texttt{ff.recip})
represent mathematical operations with only partial semantics.
Mathematically speaking, division by zero is undefined, and computing the
reciprocal of zero is undefined.
Yet, SMT-LIB requires functions to have total semantics.
We require solvers to interpret the reciprocal of zero as zero.
Moreover dividing any value by zero gives zero.
This choice is somewhat arbitrary.
It is acceptable because it easy for solvers to meet and
for verification tools to use.

A solver can meet this requirement using a preprocessing transformation.
First, it encodes division as
multiplication by the divisor's reciprocal.
Second, it encodes the reciprocal relation
$z = \texttt{(ff.recip }x\texttt{)}$
by the following (reciprocal-free) formula:%
\begin{align*}
  [(x \ne 0) \wedge (xz = 1)]\vee[(x = 0) \wedge (z = 0)]
\end{align*}
This ensures that $0$'s reciprocal is $0$.
The are other encodings of reciprocal
that do not explicitly contain a disjunction.
For example, as $(zzx=z) \land (zxx=x)$.

This requirement is also easy for verification tools that use SMT to work
with.
In particular, a verification tool can create an SMT query where division or
reciprocal have different semantics by wrapping them with an if-then-else term
that implements those semantics when the appropriate input is zero.

\begin{table}
  \centering
  \begin{tabular}{lll}
    \toprule
    Identifier & Sort & Meaning \\
    \midrule
    \texttt{ff.add} & $F \times F \to F$ & finite field addition \\
    \texttt{ff.sub} & $F \times F \to F$ & finite field subtraction \\
    \texttt{ff.mul} & $F \times F \to F$ & finite field multiplication \\
    \texttt{ff.div} & $F \times F \to F$ & finite field division \\
    \texttt{ff.neg} & $F \to F$ & finite field negation \\
    \texttt{ff.recip} & $F \to F$ & finite field reciprocal \\
    \bottomrule
  \end{tabular}
  \caption{
    A summary of finite field operations for a finite field type $F$.
  }
  \label{tab:syntax:ops}
\end{table}

\subsection{Comparison}
\label{sec:theory:eq}
Since finite fields are not ordered, the theory of finite fields only supports the equality predicate:
\begin{center}
  \texttt{(= \ffsortp \ffsortp)}\\
  \texttt{(= \ffsortpn \ffsortpn)}\\
\end{center}

\begin{example}\label{ex:ops}
  Continuing with the definition of Example~\ref{ex:define}.
  First define some variables:
  \begin{lstlisting}
  (declare-fun x0 () FF5)
  (declare-fun x1 () FF5)
  (declare-fun x2 () FF5)
  \end{lstlisting}
  Then add some assertions:
  \begin{lstlisting}
  (assert (= (ff.mul x1 x2) (ff.add x1 x2)))
  (assert (= (ff.recip x1) x0))
  (assert (= (ff.sub x2 x0)) (as ff1 FF5))
  \end{lstlisting}
  This encodes the constraint system in $\F_5$:
  \begin{equation*}
    \begin{split}
      x_1x_2 &= x_1 + x_2\\
      x_1^{-1} &= x_0\\
      x_2 - x_0 &= 1
    \end{split}
  \end{equation*}
\end{example}

\section{Existing Finite Field Solvers}\label{sec:solvers}
There are two existing SMT solvers that support the theory of finite fields:
Yices~\cite{dutertre2014yices} and cvc5~\cite{cvc5}.
Both support the logic of quantifier-free finite field arithmetic \texttt{QF\_FFA} for prime fields as defined in this paper.

\begin{itemize}
\item \textbf{Yices2} implements reasoning over prime fields using its
  MCSat engine~\cite{hader2024mcsatbased}.
  This implementation is based on the approach by Hader et al.~\cite{LPAR:HKK23}.
  Processing of polynomials over prime fields is done done using an updated version of the LibPoly library~\cite{jovanovic2017smt}.

\item \textbf{cvc5}'s prime field solver is a \cdclt theory solver
  that
  implements two decision procedures designed by Ozdemir et al.
  The first procedure is based on \gbs and triangular
  decomposition~\cite{ozdemir2023satisfiability}.
  The second is based on the same algebraic ideas,
  but uses multiple, small \gbs for better scalability in some
  cases~\cite{split_gb}.
  The implementation uses the CoCoALib computer algebra
  library~\cite{abbott2018grobner}.
\end{itemize}

\section{Future Directions}\label{sec:future}
In designing our theory,
we have intentionally omitted a number of potential features.
Some of these features might be good additions in the future.
We discuss two such features here, together with why they might be useful.

\paragraph{Conversions}
In this proposal,
we do not give operations for converting between finite field elements
and 
other discrete arithmetic types,
such as integers and bit-vectors.
This might be useful for verification problems about code that converts between
these types.
For example, the AES-GCM block cipher alternates between treating its data as
bit-vectors
and elements of $\F_{2^n}$,
to perform different kinds of operations on that data.
The bit-vector representation is used for the AES permutation
and the field representation is used in the GCM message authentication code.
Another example is an implementation of $\FF_p$ arithmetic
on a $b$-bit CPU, where $2^b \ll p$.
Such an implementation is defined by bit-vector arithmetic,
but the specification is an equation in $\FF_p$.
Thus, giving a natural statement of the implementation's correctness requires
operations to convert between $\FF_p$ and bit-vectors.
Since some SMT solvers already allow conversions between bit-vectors and
integers,
conversions between integers and finite field elements might suffice.

Another kind of conversion which might be useful is one between a field
$\FF_{p^n}$ and some extension $\FF_{(p^n)^e}$
of it (for $e > 1$).
%

\paragraph{Variable-sized fields}
In this proposal,
we consider only fields of fixed size.
This bars the possibility of queries that verify a property that holds
generically for many or all fields.
Such properties arise naturally in many verification problems.
For instance, one might have a function that implements some finite-field
operation in
which the size of the field is an input to the function.
To verify the function for all fields,
one might construct a logical formula in which the field size is a variable.

\paragraph{Acknowledgements.}
We thank
Ahmed Irfan,
Alp Bassa,
Clark Barrett,
Daniela Kaufmann,
Gereon Kremer,
Shankara Pailoor,
Sorawee Porncharoenwase,
and the
SMT'24 reviewers
for valuable discussion and feedback.
We further thank St\'ephane Graham-Lengrand for hosting the first author for a research stay at SRI during which the idea for this work initiated.
%
We acknowledge funding from the
TU Wien SecInt Doctoral College,
NSF grant number 2110397,
the Stanford Center for Automated Reasoning,
and the Simons Foundation.

\begin{flushleft}
\footnotesize
\setlength{\bibsep}{1pt}
\bibliography{main}

\begin{thebibliography}{68}
\expandafter\ifx\csname natexlab\endcsname\relax\def\natexlab#1{#1}\fi
\providecommand{\url}[1]{\texttt{#1}}
\providecommand{\href}[2]{#2}
\providecommand{\path}[1]{#1}
\providecommand{\DOIprefix}{doi:}
\providecommand{\ArXivprefix}{arXiv:}
\providecommand{\URLprefix}{URL: }
\providecommand{\Pubmedprefix}{pmid:}
\providecommand{\doi}[1]{\href{http://dx.doi.org/#1}{\path{#1}}}
\providecommand{\Pubmed}[1]{\href{pmid:#1}{\path{#1}}}
\providecommand{\bibinfo}[2]{#2}
\ifx\xfnm\relax \def\xfnm[#1]{\unskip,\space#1}\fi
\bibitem[{Barker et~al.(2018)Barker, Chen, Roginsky, Vassilev, and
  Davis}]{nist_ec}
\bibinfo{author}{E.~Barker}, \bibinfo{author}{L.~Chen},
  \bibinfo{author}{A.~Roginsky}, \bibinfo{author}{A.~Vassilev},
  \bibinfo{author}{R.~Davis}, \bibinfo{title}{Recommendation for pair-wise
  key-establishment schemes using discrete logarithm cryptography},
  \bibinfo{year}{2018}.
  \bibinfo{note}{\url{https://doi.org/10.6028/NIST.SP.800-56Ar3}}.
\bibitem[{Kotzias et~al.(2018)Kotzias, Razaghpanah, Amann, Paterson,
  Vallina-Rodriguez, and Caballero}]{longitudintaltls}
\bibinfo{author}{P.~Kotzias}, \bibinfo{author}{A.~Razaghpanah},
  \bibinfo{author}{J.~Amann}, \bibinfo{author}{K.~G. Paterson},
  \bibinfo{author}{N.~Vallina-Rodriguez}, \bibinfo{author}{J.~Caballero},
\newblock \bibinfo{title}{Coming of age: A longitudinal study of {TLS}
  deployment},
\newblock in: \bibinfo{booktitle}{IMC}, \bibinfo{year}{2018}.
\bibitem[{Anderson and McGrew(2019)}]{anderson2019tls}
\bibinfo{author}{B.~Anderson}, \bibinfo{author}{D.~McGrew},
\newblock \bibinfo{title}{{TLS} beyond the browser: Combining end host and
  network data to understand application behavior},
\newblock in: \bibinfo{booktitle}{IMC}, \bibinfo{year}{2019}.
\bibitem[{Bernstein(2005)}]{bernstein2005poly1305}
\bibinfo{author}{D.~J. Bernstein},
\newblock \bibinfo{title}{The {Poly1305}-{AES} message-authentication code},
\newblock in: \bibinfo{booktitle}{{FSE}}, \bibinfo{year}{2005}.
\bibitem[{Salowey et~al.(2008)Salowey, Choudhury, and McGrew}]{salowey2008rfc}
\bibinfo{author}{J.~Salowey}, \bibinfo{author}{A.~Choudhury},
  \bibinfo{author}{D.~McGrew}, \bibinfo{title}{Rfc 5288: {AES} galois counter
  mode ({GCM}) cipher suites for {TLS}}, \bibinfo{year}{2008}.
\bibitem[{Goldwasser et~al.(1985)Goldwasser, Micali, and
  Rackoff}]{STOC:GolMicRac85}
\bibinfo{author}{S.~Goldwasser}, \bibinfo{author}{S.~Micali},
  \bibinfo{author}{C.~Rackoff},
\newblock \bibinfo{title}{The knowledge complexity of interactive proof-systems
  (extended abstract)},
\newblock in: \bibinfo{booktitle}{STOC}, \bibinfo{year}{1985}.
\bibitem[{Parno et~al.(2016)Parno, Howell, Gentry, and
  Raykova}]{parno2016pinocchio}
\bibinfo{author}{B.~Parno}, \bibinfo{author}{J.~Howell},
  \bibinfo{author}{C.~Gentry}, \bibinfo{author}{M.~Raykova},
\newblock \bibinfo{title}{Pinocchio: Nearly practical verifiable computation},
\newblock \bibinfo{journal}{Communications of the ACM}  (\bibinfo{year}{2016}).
\bibitem[{Groth(2016)}]{groth2016size}
\bibinfo{author}{J.~Groth},
\newblock \bibinfo{title}{On the size of pairing-based non-interactive
  arguments},
\newblock in: \bibinfo{booktitle}{EUROCRYPT}, \bibinfo{year}{2016}.
\bibitem[{Chaliasos et~al.(2024)Chaliasos, Ernstberger, Theodore, Wong,
  Jahanara, and Livshits}]{zkp_vuln_sok}
\bibinfo{author}{S.~Chaliasos}, \bibinfo{author}{J.~Ernstberger},
  \bibinfo{author}{D.~Theodore}, \bibinfo{author}{D.~Wong},
  \bibinfo{author}{M.~Jahanara}, \bibinfo{author}{B.~Livshits},
\newblock \bibinfo{title}{{SoK}: What don't we know? understanding security
  vulnerabilities in {SNARKs}},
\newblock \bibinfo{journal}{arXiv preprint arXiv:2402.15293}
  (\bibinfo{year}{2024}).
\bibitem[{Damg{\aa}rd et~al.(2012)Damg{\aa}rd, Pastro, Smart, and
  Zakarias}]{damgaard2012multiparty}
\bibinfo{author}{I.~Damg{\aa}rd}, \bibinfo{author}{V.~Pastro},
  \bibinfo{author}{N.~Smart}, \bibinfo{author}{S.~Zakarias},
\newblock \bibinfo{title}{Multiparty computation from somewhat homomorphic
  encryption},
\newblock in: \bibinfo{booktitle}{CRYPTO}, \bibinfo{year}{2012}.
\bibitem[{Hastings et~al.(2019)Hastings, Hemenway, Noble, and
  Zdancewic}]{mpc_sok}
\bibinfo{author}{M.~Hastings}, \bibinfo{author}{B.~Hemenway},
  \bibinfo{author}{D.~Noble}, \bibinfo{author}{S.~Zdancewic},
\newblock \bibinfo{title}{{SoK}: General purpose compilers for secure
  multi-party computation},
\newblock in: \bibinfo{booktitle}{IEEE S\&P}, \bibinfo{year}{2019}.
\bibitem[{Regev(2009)}]{regev2009lattices}
\bibinfo{author}{O.~Regev},
\newblock \bibinfo{title}{On lattices, learning with errors, random linear
  codes, and cryptography},
\newblock \bibinfo{journal}{J.~ACM}  (\bibinfo{year}{2009}).
\bibitem[{Viand et~al.(2021)Viand, Jattke, and Hithnawi}]{fhe_sok}
\bibinfo{author}{A.~Viand}, \bibinfo{author}{P.~Jattke},
  \bibinfo{author}{A.~Hithnawi},
\newblock \bibinfo{title}{{SoK}: Fully homomorphic encryption compilers},
\newblock in: \bibinfo{booktitle}{IEEE S\&P}, \bibinfo{year}{2021}.
\bibitem[{Leino(2010)}]{leino2010dafny}
\bibinfo{author}{K.~R.~M. Leino},
\newblock \bibinfo{title}{Dafny: An automatic program verifier for functional
  correctness},
\newblock in: \bibinfo{booktitle}{LPAR}, \bibinfo{year}{2010}.
\bibitem[{Barnett et~al.(2006)Barnett, Chang, DeLine, Jacobs, and
  Leino}]{barnett2006boogie}
\bibinfo{author}{M.~Barnett}, \bibinfo{author}{B.-Y.~E. Chang},
  \bibinfo{author}{R.~DeLine}, \bibinfo{author}{B.~Jacobs},
  \bibinfo{author}{K.~R.~M. Leino},
\newblock \bibinfo{title}{Boogie: A modular reusable verifier for
  object-oriented programs},
\newblock in: \bibinfo{booktitle}{Formal Methods for Components and Objects},
  \bibinfo{year}{2006}.
\bibitem[{Dummit and Foote(2004)}]{dummit2004abstract}
\bibinfo{author}{D.~S. Dummit}, \bibinfo{author}{R.~M. Foote},
  \bibinfo{title}{Abstract algebra}, volume~\bibinfo{volume}{3},
  \bibinfo{publisher}{Wiley Hoboken}, \bibinfo{year}{2004}.
\bibitem[{Ozdemir et~al.(2023)Ozdemir, Kremer, Tinelli, and
  Barrett}]{ozdemir2023satisfiability}
\bibinfo{author}{A.~Ozdemir}, \bibinfo{author}{G.~Kremer},
  \bibinfo{author}{C.~Tinelli}, \bibinfo{author}{C.~Barrett},
\newblock \bibinfo{title}{Satisfiability modulo finite fields},
\newblock in: \bibinfo{booktitle}{CAV}, \bibinfo{year}{2023}.
\bibitem[{Niemetz et~al.(2024)Niemetz, Preiner, and Zohar}]{bb_abstractions}
\bibinfo{author}{A.~Niemetz}, \bibinfo{author}{M.~Preiner},
  \bibinfo{author}{Y.~Zohar},
\newblock \bibinfo{title}{Scalable bit-blasting with abstractions},
\newblock in: \bibinfo{booktitle}{CAV}, \bibinfo{year}{2024}.
\bibitem[{Jovanovic et~al.(2013)Jovanovic, Barrett, and
  De~Moura}]{jovanovic2013design}
\bibinfo{author}{D.~Jovanovic}, \bibinfo{author}{C.~Barrett},
  \bibinfo{author}{L.~De~Moura},
\newblock \bibinfo{title}{The design and implementation of the model
  constructing satisfiability calculus},
\newblock in: \bibinfo{booktitle}{FMCAD}, \bibinfo{year}{2013}.
\bibitem[{Hader(2022)}]{hadermsthesis}
\bibinfo{author}{T.~Hader}, \bibinfo{title}{Non-linear {SMT}-reasoning over
  finite fields}, \bibinfo{year}{2022}. \bibinfo{note}{MSc Thesis (TU Wien)}.
\bibitem[{Hader and Kovács(2022)}]{hadersmt22}
\bibinfo{author}{T.~Hader}, \bibinfo{author}{L.~Kovács},
\newblock \bibinfo{title}{Non-linear {SMT}-reasoning over finite fields},
\newblock in: \bibinfo{booktitle}{SMT}, \bibinfo{year}{2022}. \URLprefix
  \url{http://ceur-ws.org/Vol-3185/extended3245.pdf}, \bibinfo{note}{extended
  Abstract}.
\bibitem[{Hader et~al.(2023)Hader, Kaufmann, and Kov{\'{a}}cs}]{LPAR:HKK23}
\bibinfo{author}{T.~Hader}, \bibinfo{author}{D.~Kaufmann},
  \bibinfo{author}{L.~Kov{\'{a}}cs},
\newblock \bibinfo{title}{{SMT} solving over finite field arithmetic},
\newblock in: \bibinfo{booktitle}{LPAR}, \bibinfo{year}{2023}.
\bibitem[{Dutertre(2014)}]{dutertre2014yices}
\bibinfo{author}{B.~Dutertre},
\newblock \bibinfo{title}{Yices 2.2},
\newblock in: \bibinfo{booktitle}{CAV}, \bibinfo{year}{2014}.
\bibitem[{Hader et~al.(2024)Hader, Kaufmann, Irfan, Graham-Lengrand, and
  Kovács}]{hader2024mcsatbased}
\bibinfo{author}{T.~Hader}, \bibinfo{author}{D.~Kaufmann},
  \bibinfo{author}{A.~Irfan}, \bibinfo{author}{S.~Graham-Lengrand},
  \bibinfo{author}{L.~Kovács},
\newblock \bibinfo{title}{{MCSat-based Finite Field Reasoning in the Yices2 SMT
  Solver}},
\newblock in: \bibinfo{booktitle}{IJCAR}, \bibinfo{year}{2024}.
\bibitem[{Ozdemir et~al.(2024)Ozdemir, Pailoor, Bassa, Ferles, Barrett, and
  Dillig}]{split_gb}
\bibinfo{author}{A.~Ozdemir}, \bibinfo{author}{S.~Pailoor},
  \bibinfo{author}{A.~Bassa}, \bibinfo{author}{K.~Ferles},
  \bibinfo{author}{C.~Barrett}, \bibinfo{author}{I.~Dillig},
  \bibinfo{title}{Split {G}röbner bases for satisfiability modulo finite
  fields}, \bibinfo{year}{2024}. \bibinfo{note}{\url{https://ia.cr/2024/572}}.
\bibitem[{Barbosa et~al.(2022)Barbosa, Barrett, Brain, Kremer, Lachnitt, Mann,
  Mohamed, Mohamed, Niemetz, N{\"{o}}tzli, Ozdemir, Preiner, Reynolds, Sheng,
  Tinelli, and Zohar}]{cvc5}
\bibinfo{author}{H.~Barbosa}, \bibinfo{author}{C.~W. Barrett},
  \bibinfo{author}{M.~Brain}, \bibinfo{author}{G.~Kremer},
  \bibinfo{author}{H.~Lachnitt}, \bibinfo{author}{M.~Mann},
  \bibinfo{author}{A.~Mohamed}, \bibinfo{author}{M.~Mohamed},
  \bibinfo{author}{A.~Niemetz}, \bibinfo{author}{A.~N{\"{o}}tzli},
  \bibinfo{author}{A.~Ozdemir}, \bibinfo{author}{M.~Preiner},
  \bibinfo{author}{A.~Reynolds}, \bibinfo{author}{Y.~Sheng},
  \bibinfo{author}{C.~Tinelli}, \bibinfo{author}{Y.~Zohar},
\newblock \bibinfo{title}{cvc5: {A} versatile and industrial-strength {SMT}
  solver},
\newblock in: \bibinfo{booktitle}{TACAS}, \bibinfo{year}{2022}.
\bibitem[{Ozdemir et~al.(2023)Ozdemir, Wahby, Brown, and Barrett}]{ffblast}
\bibinfo{author}{A.~Ozdemir}, \bibinfo{author}{R.~S. Wahby},
  \bibinfo{author}{F.~Brown}, \bibinfo{author}{C.~Barrett},
\newblock \bibinfo{title}{Bounded verification for finite-field-blasting},
\newblock in: \bibinfo{booktitle}{CAV}, \bibinfo{year}{2023}.
\bibitem[{Pailoor et~al.(2023)Pailoor, Chen, Wang, Rodr{\'\i}guez, Van~Geffen,
  Morton, Chu, Gu, Feng, and Dillig}]{PLDI:PCWRVMCGFD23}
\bibinfo{author}{S.~Pailoor}, \bibinfo{author}{Y.~Chen},
  \bibinfo{author}{F.~Wang}, \bibinfo{author}{C.~Rodr{\'\i}guez},
  \bibinfo{author}{J.~Van~Geffen}, \bibinfo{author}{J.~Morton},
  \bibinfo{author}{M.~Chu}, \bibinfo{author}{B.~Gu}, \bibinfo{author}{Y.~Feng},
  \bibinfo{author}{I.~Dillig},
\newblock \bibinfo{title}{Automated detection of under-constrained circuits in
  zero-knowledge proofs},
\newblock in: \bibinfo{booktitle}{PLDI}, \bibinfo{year}{2023}.
\bibitem[{Bell{\'e}s-Mu{\~n}oz et~al.(2022)Bell{\'e}s-Mu{\~n}oz, Isabel,
  Mu{\~n}oz-Tapia, Rubio, and Baylina}]{circom}
\bibinfo{author}{M.~Bell{\'e}s-Mu{\~n}oz}, \bibinfo{author}{M.~Isabel},
  \bibinfo{author}{J.~L. Mu{\~n}oz-Tapia}, \bibinfo{author}{A.~Rubio},
  \bibinfo{author}{J.~Baylina},
\newblock \bibinfo{title}{Circom: A circuit description language for building
  zero-knowledge applications},
\newblock \bibinfo{journal}{IEEE Transactions on Dependable and Secure
  Computing}  (\bibinfo{year}{2022}).
\bibitem[{Soureshjani et~al.(2023)Soureshjani, Hall-Andersen, Jahanara, Kam,
  Gorzny, and Ahmadvand}]{halo2ver}
\bibinfo{author}{F.~H. Soureshjani}, \bibinfo{author}{M.~Hall-Andersen},
  \bibinfo{author}{M.~Jahanara}, \bibinfo{author}{J.~Kam},
  \bibinfo{author}{J.~Gorzny}, \bibinfo{author}{M.~Ahmadvand},
  \bibinfo{title}{Automated analysis of {Halo2} circuits},
  \bibinfo{year}{2023}. \bibinfo{note}{\url{https://ia.cr/2023/1051}}.
\bibitem[{Zero(2022)}]{plonky2}
\bibinfo{author}{P.~Zero}, \bibinfo{title}{Plonky2: Fast recursive arguments
  with {Plonk} and {FRI}}, \bibinfo{year}{2022}.
  \bibinfo{note}{\url{https://github.com/0xPolygonZero/plonky2/blob/main/plonky2/plonky2.pdf}}.
\bibitem[{Ben-Sasson et~al.(2018)Ben-Sasson, Bentov, Horesh, and
  Riabzev}]{ben2018fast}
\bibinfo{author}{E.~Ben-Sasson}, \bibinfo{author}{I.~Bentov},
  \bibinfo{author}{Y.~Horesh}, \bibinfo{author}{M.~Riabzev},
\newblock \bibinfo{title}{Fast reed-solomon interactive oracle proofs of
  proximity},
\newblock in: \bibinfo{booktitle}{ICALP}, \bibinfo{year}{2018}.
\bibitem[{Erbsen et~al.(2020)Erbsen, Philipoom, Gross, Sloan, and
  Chlipala}]{fiat_crypto}
\bibinfo{author}{A.~Erbsen}, \bibinfo{author}{J.~Philipoom},
  \bibinfo{author}{J.~Gross}, \bibinfo{author}{R.~Sloan},
  \bibinfo{author}{A.~Chlipala},
\newblock \bibinfo{title}{Simple high-level code for cryptographic arithmetic
  -- with proofs, without compromises}  (\bibinfo{year}{2020}).
\bibitem[{Barthe et~al.(2012)Barthe, Dupressoir, Gr{\'e}goire, Kunz, Schmidt,
  and Strub}]{barthe2012easycrypt}
\bibinfo{author}{G.~Barthe}, \bibinfo{author}{F.~Dupressoir},
  \bibinfo{author}{B.~Gr{\'e}goire}, \bibinfo{author}{C.~Kunz},
  \bibinfo{author}{B.~Schmidt}, \bibinfo{author}{P.-Y. Strub},
\newblock \bibinfo{title}{Easycrypt: A tutorial},
\newblock \bibinfo{journal}{International School on Foundations of Security
  Analysis and Design}  (\bibinfo{year}{2012}) \bibinfo{pages}{146--166}.
\bibitem[{Zinzindohou{\'e} et~al.(2017)Zinzindohou{\'e}, Bhargavan, Protzenko,
  and Beurdouche}]{zinzindohoue2017hacl}
\bibinfo{author}{J.-K. Zinzindohou{\'e}}, \bibinfo{author}{K.~Bhargavan},
  \bibinfo{author}{J.~Protzenko}, \bibinfo{author}{B.~Beurdouche},
\newblock \bibinfo{title}{{HACL*}: A verified modern cryptographic library},
\newblock in: \bibinfo{booktitle}{CCS}, \bibinfo{year}{2017}.
\bibitem[{Almeida et~al.(2017)Almeida, Barbosa, Barthe, Blot, Gr{\'e}goire,
  Laporte, Oliveira, Pacheco, Schmidt, and Strub}]{almeida2017jasmin}
\bibinfo{author}{J.~B. Almeida}, \bibinfo{author}{M.~Barbosa},
  \bibinfo{author}{G.~Barthe}, \bibinfo{author}{A.~Blot},
  \bibinfo{author}{B.~Gr{\'e}goire}, \bibinfo{author}{V.~Laporte},
  \bibinfo{author}{T.~Oliveira}, \bibinfo{author}{H.~Pacheco},
  \bibinfo{author}{B.~Schmidt}, \bibinfo{author}{P.-Y. Strub},
\newblock \bibinfo{title}{Jasmin: High-assurance and high-speed cryptography},
\newblock in: \bibinfo{booktitle}{CCS}, \bibinfo{year}{2017}.
\bibitem[{Coglio et~al.(2023)Coglio, McCarthy, and Smith}]{kestrel}
\bibinfo{author}{A.~Coglio}, \bibinfo{author}{E.~McCarthy},
  \bibinfo{author}{E.~W. Smith}, \bibinfo{title}{Formal verification of
  zero-knowledge circuits}, \bibinfo{year}{2023}. \URLprefix
  \url{http://dx.doi.org/10.4204/EPTCS.393.9}.
\bibitem[{Liu et~al.(2023)Liu, Kretz, Liu, Tan, Wang, Sun, Pearson, Miltner,
  Dillig, and Feng}]{liu2023certifying}
\bibinfo{author}{J.~Liu}, \bibinfo{author}{I.~Kretz}, \bibinfo{author}{H.~Liu},
  \bibinfo{author}{B.~Tan}, \bibinfo{author}{J.~Wang},
  \bibinfo{author}{Y.~Sun}, \bibinfo{author}{L.~Pearson},
  \bibinfo{author}{A.~Miltner}, \bibinfo{author}{I.~Dillig},
  \bibinfo{author}{Y.~Feng},
\newblock \bibinfo{title}{Certifying zero-knowledge circuits with refinement
  types},
\newblock \bibinfo{journal}{arXiv preprint arXiv:2304.07648}
  (\bibinfo{year}{2023}).
\bibitem[{Jiang et~al.(2023)Jiang, Chait-Roth, DeStefano, Walfish, and
  Wies}]{jiang2023less}
\bibinfo{author}{K.~Jiang}, \bibinfo{author}{D.~Chait-Roth},
  \bibinfo{author}{Z.~DeStefano}, \bibinfo{author}{M.~Walfish},
  \bibinfo{author}{T.~Wies},
\newblock \bibinfo{title}{Less is more: refinement proofs for probabilistic
  proofs},
\newblock in: \bibinfo{booktitle}{IEEE S\&P}, \bibinfo{year}{2023}.
\bibitem[{Ekici et~al.(2017)Ekici, Mebsout, Tinelli, Keller, Katz, Reynolds,
  and Barrett}]{ekici2017smtcoq}
\bibinfo{author}{B.~Ekici}, \bibinfo{author}{A.~Mebsout},
  \bibinfo{author}{C.~Tinelli}, \bibinfo{author}{C.~Keller},
  \bibinfo{author}{G.~Katz}, \bibinfo{author}{A.~Reynolds},
  \bibinfo{author}{C.~Barrett},
\newblock \bibinfo{title}{{SMTCoq}: A plug-in for integrating {SMT} solvers
  into {Coq}},
\newblock in: \bibinfo{booktitle}{CAV}, \bibinfo{year}{2017}.
\bibitem[{Meier et~al.(2013)Meier, Schmidt, Cremers, and
  Basin}]{meier2013tamarin}
\bibinfo{author}{S.~Meier}, \bibinfo{author}{B.~Schmidt},
  \bibinfo{author}{C.~Cremers}, \bibinfo{author}{D.~Basin},
\newblock \bibinfo{title}{The {TAMARIN} prover for the symbolic analysis of
  security protocols},
\newblock in: \bibinfo{booktitle}{Computer Aided Verification: 25th
  International Conference, CAV 2013, Saint Petersburg, Russia, July 13-19,
  2013. Proceedings 25}, \bibinfo{organization}{Springer},
  \bibinfo{year}{2013}, pp. \bibinfo{pages}{696--701}.
\bibitem[{Blanchet et~al.(2018)Blanchet, Smyth, Cheval, and
  Sylvestre}]{blanchet2018proverif}
\bibinfo{author}{B.~Blanchet}, \bibinfo{author}{B.~Smyth},
  \bibinfo{author}{V.~Cheval}, \bibinfo{author}{M.~Sylvestre},
\newblock \bibinfo{title}{{ProVerif} 2.00: automatic cryptographic protocol
  verifier, user manual and tutorial}  (\bibinfo{year}{2018}).
\bibitem[{Cohn-Gordon et~al.(2020)Cohn-Gordon, Cremers, Dowling, Garratt, and
  Stebila}]{cohn2020formal}
\bibinfo{author}{K.~Cohn-Gordon}, \bibinfo{author}{C.~Cremers},
  \bibinfo{author}{B.~Dowling}, \bibinfo{author}{L.~Garratt},
  \bibinfo{author}{D.~Stebila},
\newblock \bibinfo{title}{A formal security analysis of the {S}ignal messaging
  protocol},
\newblock \bibinfo{journal}{Journal of Cryptology} \bibinfo{volume}{33}
  (\bibinfo{year}{2020}) \bibinfo{pages}{1914--1983}.
\bibitem[{Cremers and Jackson(2019)}]{cremers2019prime}
\bibinfo{author}{C.~Cremers}, \bibinfo{author}{D.~Jackson},
\newblock \bibinfo{title}{Prime, order please! revisiting small subgroup and
  invalid curve attacks on protocols using {D}iffie-{H}ellman},
\newblock in: \bibinfo{booktitle}{CSF}, \bibinfo{year}{2019}.
\bibitem[{{\'A}brah{\'a}m et~al.(2021){\'A}brah{\'a}m, Davenport, England, and
  Kremer}]{abraham2021deciding}
\bibinfo{author}{E.~{\'A}brah{\'a}m}, \bibinfo{author}{J.~H. Davenport},
  \bibinfo{author}{M.~England}, \bibinfo{author}{G.~Kremer},
\newblock \bibinfo{title}{Deciding the consistency of non-linear real
  arithmetic constraints with a conflict driven search using cylindrical
  algebraic coverings},
\newblock \bibinfo{journal}{Journal of Logical and Algebraic Methods in
  Programming} \bibinfo{volume}{119} (\bibinfo{year}{2021}).
\bibitem[{Cimatti et~al.(2018)Cimatti, Griggio, Irfan, Roveri, and
  Sebastiani}]{cimatti2018experimenting}
\bibinfo{author}{A.~Cimatti}, \bibinfo{author}{A.~Griggio},
  \bibinfo{author}{A.~Irfan}, \bibinfo{author}{M.~Roveri},
  \bibinfo{author}{R.~Sebastiani},
\newblock \bibinfo{title}{Experimenting on solving nonlinear integer arithmetic
  with incremental linearization},
\newblock in: \bibinfo{booktitle}{International Conference on Theory and
  Applications of Satisfiability Testing}, \bibinfo{organization}{Springer},
  \bibinfo{year}{2018}, pp. \bibinfo{pages}{383--398}.
\bibitem[{Mar{\'e}chal et~al.(2016)Mar{\'e}chal, Fouilh{\'e}, King, Monniaux,
  and P{\'e}rin}]{marechal2016polyhedral}
\bibinfo{author}{A.~Mar{\'e}chal}, \bibinfo{author}{A.~Fouilh{\'e}},
  \bibinfo{author}{T.~King}, \bibinfo{author}{D.~Monniaux},
  \bibinfo{author}{M.~P{\'e}rin},
\newblock \bibinfo{title}{Polyhedral approximation of multivariate polynomials
  using {H}andelman’s theorem},
\newblock in: \bibinfo{booktitle}{VMCAI}, \bibinfo{year}{2016}.
\bibitem[{Fr{\"a}nzle et~al.(2006)Fr{\"a}nzle, Herde, Teige, Ratschan, and
  Schubert}]{franzle2006efficient}
\bibinfo{author}{M.~Fr{\"a}nzle}, \bibinfo{author}{C.~Herde},
  \bibinfo{author}{T.~Teige}, \bibinfo{author}{S.~Ratschan},
  \bibinfo{author}{T.~Schubert},
\newblock \bibinfo{title}{Efficient solving of large non-linear arithmetic
  constraint systems with complex boolean structure},
\newblock \bibinfo{journal}{Journal on Satisfiability, Boolean Modeling and
  Computation} \bibinfo{volume}{1} (\bibinfo{year}{2006}).
\bibitem[{Tung et~al.(2016)Tung, Khanh, and Ogawa}]{tung2016rasat}
\bibinfo{author}{V.~X. Tung}, \bibinfo{author}{T.~V. Khanh},
  \bibinfo{author}{M.~Ogawa},
\newblock \bibinfo{title}{{raSAT}: An {SMT} solver for polynomial constraints},
\newblock in: \bibinfo{booktitle}{IJCAR}, \bibinfo{year}{2016}.
\bibitem[{Jovanovi{\'c} and Moura(2011)}]{jovanovic2011cutting}
\bibinfo{author}{D.~Jovanovi{\'c}}, \bibinfo{author}{L.~d. Moura},
\newblock \bibinfo{title}{Cutting to the chase solving linear integer
  arithmetic},
\newblock in: \bibinfo{booktitle}{CADE}, \bibinfo{year}{2011}.
\bibitem[{Dillig et~al.(2009)Dillig, Dillig, and Aiken}]{dillig2009cuts}
\bibinfo{author}{I.~Dillig}, \bibinfo{author}{T.~Dillig},
  \bibinfo{author}{A.~Aiken},
\newblock \bibinfo{title}{Cuts from proofs: A complete and practical technique
  for solving linear inequalities over integers},
\newblock \bibinfo{year}{2009}, pp. \bibinfo{pages}{233--247}.
\bibitem[{Corzilius et~al.(2015)Corzilius, Kremer, Junges, Schupp, and
  {\'A}brah{\'a}m}]{corzilius2015smt}
\bibinfo{author}{F.~Corzilius}, \bibinfo{author}{G.~Kremer},
  \bibinfo{author}{S.~Junges}, \bibinfo{author}{S.~Schupp},
  \bibinfo{author}{E.~{\'A}brah{\'a}m},
\newblock \bibinfo{title}{{SMT}-{RAT}: an open source {C++} toolbox for
  strategic and parallel {SMT} solving},
\newblock in: \bibinfo{booktitle}{SAT}, \bibinfo{year}{2015}.
\bibitem[{Jovanovi{\'c} and De~Moura(2013)}]{jovanovic2013solving}
\bibinfo{author}{D.~Jovanovi{\'c}}, \bibinfo{author}{L.~De~Moura},
\newblock \bibinfo{title}{Solving non-linear arithmetic},
\newblock \bibinfo{journal}{{ACM} Communications in Computer Algebra}
  \bibinfo{volume}{46} (\bibinfo{year}{2013}).
\bibitem[{Jovanovi{\'c}(2017)}]{jovanovic2017solving}
\bibinfo{author}{D.~Jovanovi{\'c}},
\newblock \bibinfo{title}{Solving nonlinear integer arithmetic with {MCSAT}},
\newblock in: \bibinfo{booktitle}{VMCAI}, \bibinfo{year}{2017}.
\bibitem[{Moura and Jovanovi{\'c}(2013)}]{moura2013model}
\bibinfo{author}{L.~d. Moura}, \bibinfo{author}{D.~Jovanovi{\'c}},
\newblock \bibinfo{title}{A model-constructing satisfiability calculus},
\newblock in: \bibinfo{booktitle}{VMCAI}, \bibinfo{year}{2013}.
\bibitem[{Cai et~al.(2023)Cai, Li, and Zhang}]{cai2023local}
\bibinfo{author}{S.~Cai}, \bibinfo{author}{B.~Li}, \bibinfo{author}{X.~Zhang},
\newblock \bibinfo{title}{Local search for satisfiability modulo integer
  arithmetic theories},
\newblock \bibinfo{journal}{{ACM} Transactions on Computational Logic}
  (\bibinfo{year}{2023}).
\bibitem[{Zhang et~al.(2024)Zhang, Li, and Cai}]{zhang2024deep}
\bibinfo{author}{X.~Zhang}, \bibinfo{author}{B.~Li}, \bibinfo{author}{S.~Cai},
\newblock \bibinfo{title}{Deep combination of {CDCL} ({T}) and local search for
  satisfiability modulo non-linear integer arithmetic theory},
\newblock \bibinfo{year}{2024}.
\bibitem[{Wang et~al.(2023)Wang, Zhan, Li, and Cai}]{wang2023efficient}
\bibinfo{author}{Z.~Wang}, \bibinfo{author}{B.~Zhan}, \bibinfo{author}{B.~Li},
  \bibinfo{author}{S.~Cai},
\newblock \bibinfo{title}{Efficient local search for nonlinear real
  arithmetic},
\newblock in: \bibinfo{booktitle}{VMCAI}, \bibinfo{year}{2023}.
\bibitem[{Caviness and Johnson(1998)}]{caviness2012quantifier}
\bibinfo{author}{B.~F. Caviness}, \bibinfo{author}{J.~R. Johnson},
  \bibinfo{title}{Quantifier elimination and cylindrical algebraic
  decomposition}, \bibinfo{publisher}{Springer Science \& Business Media},
  \bibinfo{year}{1998}.
\bibitem[{Weispfenning(1997)}]{weispfenning1997quantifier}
\bibinfo{author}{V.~Weispfenning},
\newblock \bibinfo{title}{Quantifier elimination for real algebra—the
  quadratic case and beyond},
\newblock \bibinfo{journal}{Applicable Algebra in Engineering, Communication
  and Computing} \bibinfo{volume}{8} (\bibinfo{year}{1997}).
\bibitem[{Lidl and Niederreiter(1994)}]{lidl1994introff}
\bibinfo{author}{R.~Lidl}, \bibinfo{author}{H.~Niederreiter},
  \bibinfo{title}{Introduction to finite fields and their applications},
  \bibinfo{publisher}{Cambridge university press}, \bibinfo{year}{1994}.
\bibitem[{McEliece(2012)}]{mceliece2012finite}
\bibinfo{author}{R.~J. McEliece}, \bibinfo{title}{Finite fields for computer
  scientists and engineers}, volume~\bibinfo{volume}{23},
  \bibinfo{publisher}{Springer Science \& Business Media},
  \bibinfo{year}{2012}.
\bibitem[{Gallian(2021)}]{contemporaryAbstractAlgebra}
\bibinfo{author}{J.~A. Gallian}, \bibinfo{title}{{Contemporary Abstract
  Algebra}}, \bibinfo{publisher}{Chapman and Hall/CRC}, \bibinfo{year}{2021}.
\bibitem[{Heath and Loehr(2004)}]{heath2004conway}
\bibinfo{author}{L.~S. Heath}, \bibinfo{author}{N.~A. Loehr},
\newblock \bibinfo{title}{New algorithms for generating conway polynomials over
  finite fields},
\newblock \bibinfo{journal}{Journal of Symbolic Computation}
  \bibinfo{volume}{38} (\bibinfo{year}{2004}) \bibinfo{pages}{1003--1024}.
\bibitem[{Lübeck(????)}]{conway_poly_db}
\bibinfo{author}{F.~Lübeck}, \bibinfo{title}{Conway polynomials for finite
  fields}, ???? \bibinfo{note}{Pre-computed Conway polynomials. Available at
  \url{https://www.math.rwth-aachen.de/~Frank.Luebeck/data/ConwayPol/index.html}
  or \url{https://github.com/sagemath/conway-polynomials}}.
\bibitem[{Rabin(1980)}]{rabin1980probabilistic}
\bibinfo{author}{M.~O. Rabin},
\newblock \bibinfo{title}{Probabilistic algorithm for testing primality},
\newblock \bibinfo{journal}{Journal of number theory} \bibinfo{volume}{12}
  (\bibinfo{year}{1980}) \bibinfo{pages}{128--138}.
\bibitem[{Jovanovic and Dutertre(2017)}]{jovanovic2017smt}
\bibinfo{author}{D.~Jovanovic}, \bibinfo{author}{B.~Dutertre},
\newblock \bibinfo{title}{Libpoly: {A} library for reasoning about
  polynomials},
\newblock in: \bibinfo{booktitle}{Intl. Workshop on Satisfiability Modulo
  Theories {(SMT)}}, {CEUR} Workshop Proceedings, \bibinfo{year}{2017}.
\bibitem[{Abbott and Bigatti(2018)}]{abbott2018grobner}
\bibinfo{author}{J.~Abbott}, \bibinfo{author}{A.~M. Bigatti},
\newblock \bibinfo{title}{Gr{\"o}bner bases for everyone with {CoCoA-5} and
  {CoCoALib}},
\newblock in: \bibinfo{booktitle}{The 50th Anniversary of Gr{\"o}bner Bases},
  volume~\bibinfo{volume}{77}, \bibinfo{publisher}{Mathematical Society of
  Japan}, \bibinfo{year}{2018}, pp. \bibinfo{pages}{1--25}.

\end{thebibliography}
\end{flushleft}

\appendix

\end{document}